\documentclass[aps,prd,onecolumn,groupedaddress,showpacs,nofootinbib,amssymb]{revtex4}
\usepackage[T1]{fontenc}
\usepackage[latin1]{inputenc}
\usepackage{graphicx}
\usepackage[english]{babel}
\usepackage{amsmath}
\usepackage{amssymb}
\usepackage{amsfonts}
\usepackage{bm}

\begin{document}

\def\pp{{\, \mid \hskip -1.5mm =}}
\def\cL{{\cal L}}
\def\be{\begin{equation}}
\def\ee{\end{equation}}
\def\bea{\begin{eqnarray}}
\def\eea{\end{eqnarray}}
\def\beq{\begin{eqnarray}}
\def\eeq{\end{eqnarray}}
\def\tr{{\rm tr}\, }
\def\nn{\nonumber \\}
\def\e{{\rm e}}

\title{Gauss-Bonnet Chameleon Mechanism of Dark Energy} 

\author{Yusaku Ito and Shin'ichi Nojiri}
\affiliation{Department of Physics, Nagoya University, Nagoya 464-8602, Japan}

\begin{abstract}

As a model of the current accelerated expansion of the universe, we consider 
a model of the scalar-Einstein-Gauss-Bonnet gravity. 
This model includes the propagating scalar modes, which might give a large correction 
to the Newton law. In order to avoid this problem, 
we propose an extension of the Chameleon mechanism where the scalar mode becomes massive due 
to the coupling with the Gauss-Bonnet term. Since the Gauss-Bonnet invariant does not vanish 
near the earth or in the Solar System, even in the vacuum, the scalar mode is massive even 
in the vacuum and the correction to the Newton law could be small. 
We also discuss about the possibility that the model could describe simultaneously 
the inflation in the early universe, in addition to the current accelerated expansion. 

\end{abstract}

\pacs{95.36.+x, 98.80.Cq}

\maketitle

\section{Introduction \label{SecI}}

We now believe the accelerated expansion of the current universe \cite{WMAP1, Komatsu:2008hk, SN1} 
(for recent reviews, see~\cite{Peebles:2002gy, Padmanabhan:2002ji, Copeland:2006wr, review, 
Nojiri:2008nk}). 
One scenario to explain the accelerated expansion is to introduce unknown matter/energy 
called dark energy. 
Another scenario is to modify the Einstein gravity. 
As a scenario of modified gravities, there have been proposed many kinds of models, 
like $F(R)$-gravity (for review, \cite{review, Nojiri:2008nk}), 
and the scalar-Einstein-Gauss-Bonnet gravity \cite{scalar-Gauss-Bonnet}. 
There are other many kinds of scenarios, like a non-linear higher-derivative one~\cite{Li:2005fm},
phantom coupled to dark matter with an appropriate coupling~\cite{Nojiri:2005sx},
the thermodynamical inhomogeneous dark energy model~\cite{Nojiri:2005sr},
multiple kinetic k-essence~\cite{multiple kinetic k-essence},
multi-field models (two scalar fields model~\cite{eli, Caldwell:2005ai, Wei:2005si}, 
``quintom'' consisting of phantom and canonical scalar fields~\cite{quintom}), 
and the description of those models through
the Parameterized Post-Friedmann approach~\cite{Fang:2008sn},
a classical Dirac field~\cite{Cataldo:2007vt}, 
string-inspired models~\cite{string-inspired models},
non-local gravity~\cite{Jhingan:2008ym, Deser:2007jk},
and a model in loop quantum cosmology~\cite{Singh:2005km}
(for a detailed review, see~\cite{Copeland:2006wr}). 

Many of these models commonly include the propagating 
scalar modes, which might give a large correction to the Newton law. In order to avoid this problem, 
a scenario called Chameleon mechanism has been proposed \cite{Chame}. In the scenario, 
the mass of the scalar mode 
becomes large due to the coupling with matter or scalar curvature in the Solar System or on and/or 
in the earth and the correction to the Newton law becomes very small and cannot be observed. 
The Chameleon mechanism has been used to obtain realistic models of $F(R)$-gravity 
\cite{Hu:2007nk} (for some related models, see~\cite{acc, acc1}). 
In this paper, we propose a model where the scalar mode becomes massive due to the coupling with the 
Gauss-Bonnet term. In the previous scenarios, where the scalar mode becomes massive due to 
the coupling with matter or the scalar curvature, if we observe the scalar mode in the vacuum chamber, 
where any matter does not exist inside and therefore the scalar curvature vanishes, the mass of the scalar 
mode becomes very small and the correction to the Newton law could be observed, even on the earth. 
We should note that the scalar curvature and the Ricci tensor vanish in the vacuum but the Riemann 
tensor and therefore the Gauss-Bonnet invariant do not vanish near the earth or in the Solar System, 
even in the vacuum. Therefore in the scenario proposed in this note, the scalar mode is massive even 
in the vacuum and the correction to the Newton law could be small. 

In the model proposed in this paper, there is a de Sitter solution, where the effective 
cosmological constant could be the order of the dark energy density observed in the 
present universe. In order to generate such a very small effective cosmological constant, 
which is the order of the square of the present Hubble constant, we need not so small 
mass scale in the action. The action only contains the scales of the order of the Planck 
scale and the elementary particle scale with the order of $10^4$\,GeV. We also show that 
the de Sitter solution is semi-stable, that is, the model has an instability of 
only the order much larger than that of the age of the universe. 

If we properly choose the parameters in the action, there appear two de Sitter space 
solutions. One may correspond to the current acceleratedly expanding universe. The Hubble rate 
of another solution can be the order of the Planck scale and very large. 
The solution with the large Hubble rate could be identified to describe the inflation. 
The solution is, however, stable. Then in order to make an exit from the inflation, 
we may add a small term given by another scalar field coupled with the scalar curvature, 
which could be equivalent to the non-local action \cite{Jhingan:2008ym, Deser:2007jk} and generates 
the instability of the de Sitter solution. The added term is relevant only in the 
epoch of the inflation but irrelevant to the present accelerating universe. 

\section{Model of Accelerated Expansion and Compton Length of Scalar Field \label{SecII}}

We start with the following general action of the Gauss-Bonnet gravity coupled with 
scalar field $\phi$:
\begin{equation}
\label{e1}
S=\int d^4x\,\sqrt{-g}\left[\frac{\mathcal{R}}{2\kappa^2} 
 -\frac{1}{2}\partial_\mu\phi\,\partial^\mu\phi-V(\phi)-f(\phi)\,\mathcal{R}_\mathrm{GB}\!^2\right]
 - \int d^4x\,\mathcal{L}_\mathrm{matter} 
\left(\psi_\mathrm{matter}^{(i)}, g^{(i)}_{\mu\nu}\right) \ .
\end{equation}
Here $\mathcal{R}_\mathrm{GB}$ is the Gauss-Bonnet invariant defined by 
\be
\label{e1-2}
\mathcal{R}_\mathrm{GB}\!^2 = \mathcal{R}^2 - 4 \mathcal{R}_{\mu\nu}\mathcal{R}^{\mu\nu} 
+ \mathcal{R}_{\mu\nu\rho\sigma}\mathcal{R}^{\mu\nu\rho\sigma}\ .
\ee
In (\ref{e1}), $\kappa$ is the gravitational constant, $V(\phi)$ is the potential 
of the scalar field $\phi$, and $f(\phi)$ is an appropriate function of the scalar field, 
which gives the coupling of the scalar field $\phi$ with the Gauss-Bonnet invariant. 
The matter fields are denoted by $\psi_\mathrm{matter}^{(i)}$ and we assume the scalar 
field $\phi$ couples with the matter via metric tensor: 
$g^{(i)}_{\mu\nu}\equiv \e^{2\beta_i \phi/M_\mathrm{Pl}}g_{\mu\nu}$. 
We should note that 
$\mathcal{L}_\mathrm{matter} \left(\psi_\mathrm{matter}^{(i)}, g^{(i)}_{\mu\nu}\right)$ 
is the matter Lagrangian density (pseudo scalar). 

By the variation of the scalar field, we obtain 
\be
\label{se1}
\nabla_\mu \nabla^\mu \phi = V'(\phi) + f'(\phi) \mathcal{R}_\mathrm{GB}\!^2 
+ \sum_i \frac{\beta_i}{M_\mathrm{Pl}}
\e^{2\beta_i \phi/M_\mathrm{Pl}}
g_{\mu\nu} T^{(i)\mu\nu}\ .
\ee
Here
\be
\label{se2}
T^{(i)\mu\nu}=\frac{2}{\sqrt{-g}}
\frac{\partial \mathcal{L}}{\partial g^{(i)}_{\mu\nu}}\e^{2\beta_i \phi/M_\mathrm{Pl}}\ . 
\ee
On the other hand, the variation of the metric tensor gives
\bea
\label{se3}
0&=& - \frac{M_\mathrm{Pl}\!^2}{2}\left( \mathcal{R}_{\mu\nu} - \frac{1}{2} \mathcal{R} g_{\mu\nu}\right) 
+ \frac{1}{2} \nabla_\mu \phi \nabla_\nu \phi 
 - \frac{1}{4}g_{\mu\nu} \left( \nabla_\rho \phi \nabla^\rho \phi + V(\phi) \right) 
 - 2 \left[ \nabla_\mu \nabla_\nu f(\phi) \right] \mathcal{R} \nn
&& + 4 \left[ \nabla_\mu \nabla_\rho f(\phi) \right] \mathcal{R}_\nu^{\ \rho} 
+ 4 \left[ \nabla_\nu \nabla_\rho f(\phi) \right] \mathcal{R}_\mu^{\ \rho} 
 - \left[ \nabla_\rho \nabla^\rho f(\phi) \right] \left(4 \mathcal{R}_{\mu\nu} - 2 \mathcal{R} g_{\mu\nu} \right) \nn 
&& -4 \left[ \nabla_\rho \nabla_\sigma f(\phi)\right] \left(\mathcal{R}^{\rho\sigma} g_{\mu\nu} 
 - \mathcal{R}_{\mu\ \nu\ }^{\ \rho\ \sigma} \right)
+ T^{(i)}_{\mu\nu}\ .
\eea
Especially if $\phi$ is constant and the spacetime is de Sitter space, where curvatures are 
(covariantly) constant, and if we neglect the contribution from the matter, Eqs(\ref{se1}) 
and (\ref{se3}) has the following form: 
\bea
\label{se1b}
0 &=& V'(\phi) + f'(\phi) \mathcal{R}_\mathrm{GB}\!^2 \ , \\
\label{se3b}
0 &=& - \frac{M_\mathrm{Pl}\!^2}{2}\left( \mathcal{R}_{\mu\nu} - \frac{1}{2} \mathcal{R} g_{\mu\nu}\right) 
 - \frac{1}{4}g_{\mu\nu} V(\phi) \ .
\eea
The second equation (\ref{se3b}) is not changed from the Einstein equation since 
$f(\phi) \mathcal{R}_\mathrm{GB}\!^2$ term becomes total derivative for constant $\phi$. 

In order to explain the idea of the Chameleon mechanism in the scalar-Einstein-Gauss-Bonnet gravity, 
We now consider the following model as an instructive example:
\begin{equation}
\label{e2}
V(\phi)=v_0\,M_\mathrm{Pl}\!^{4+n}\,\phi^{-n}\,\mathrm{e}^{-\alpha\phi/M_\mathrm{Pl}}\ ,\quad
f(\phi)=f_0\,\mathrm{e}^{\alpha\phi/M_\mathrm{Pl}}\ .
\end{equation}
Here $M_\mathrm{Pl}$ is the Planck scale and $\alpha$, $n$, $v_0$, $f_0$ are positive 
dimensionless parameters. 
Then in the model (\ref{e2}), there is only one dimensional parameter, the Planck scale $M_\mathrm{Pl}$. 
As we will see later, however, we need to choose $\alpha>10^{15}$ so that the correction to 
the Newton law could be small. 
We may regard $L \equiv M_\mathrm{Pl}/\alpha$, which appears in the exponential function 
in (\ref{e2}), with another scale. We find $L =M_\mathrm{Pl}/\alpha \sim 10^4\,\mathrm{GeV}$, 
which could correspond to the scale of particle physics and the scale is not so small compared with 
the typical scale of the current accelerated expansion. 
Although the model (\ref{e2}) does not contain so small parameter, the very small scale 
corresponding to the value of the present Hubble rate $H_0 \sim 10^{-33}\,\mathrm{eV}$ 
can be generated dynamically. 
We should note that the model (\ref{e2}) is not the completely consistent model as we will see later 
but it could be enough to explain the idea of the Gauss-Bonnet Chameleon mechanism. 

In the following, for the usefulness of the calculations, we define the following dimensionless 
variables: 
\begin{equation}
\label{e3}
\varphi\equiv\phi/M_\mathrm{Pl}\ , \quad
h\equiv H/M_\mathrm{Pl}\ , \quad
\mathcal{\tilde R}_\mathrm{GB}\!^2 \equiv \mathcal{R}_\mathrm{GB}\!^2/M_\mathrm{Pl}\!^4\ .
\end{equation}
By using these variables, the equation corresponding to the FRW equation, which is $(\mu,\nu)=(0,0)$ 
component of (\ref{se3}), is given by
\begin{equation}
\label{e4}
3=\frac{1}{2}\,\varphi'{^2}+v_0\,h^{-2}\,\varphi^{-n}\,
\mathrm{e}^{-\alpha\varphi}+24\,f_0\,\alpha\,h^2\,\varphi'\,\mathrm{e}^{\alpha\varphi}\ .
\end{equation}
On the other hand, the equation for the scalar field (\ref{se1}) is given by
\begin{equation}
\label{e5}
0=\varphi''+\left(3+\frac{h'}{h}\right)\varphi'-v_0\,\alpha\,h^{-2}
\left(1+\frac{n}{\alpha\varphi}\right)\varphi^{-n}\,\mathrm{e}^{-\alpha\varphi}
+24\,f_0\,\alpha\left(1+\frac{h'}{h}\right)h^2\,\mathrm{e}^{\alpha\varphi}\ .
\end{equation}
Here we denote the derivative with respect to $\ln a$ by $'$ as 
$ d/ d(\ln a)=H^{-1}\, d/ dt$. 

We now investigate if we can have a de Sitter space solution, corresponding to the 
present accelerating universe by solving (\ref{e4}) and (\ref{e5}). 
Since we have $h'=0$ in de Sitter space, by assuming the scalar field to be a constant: 
$\varphi'=0$, Eq.(\ref{e4}) has the following form: 
\begin{equation}
\label{e6}
3=v_0\,h^{-2}\,\varphi^{-n}\,\mathrm{e}^{-\alpha\varphi}\ .
\end{equation}
On the other hand, Eq.(\ref{e5}) is reduced to be
\begin{equation}
\label{e7}
0=-v_0\,h^{-2}\left(1+\frac{n}{\alpha\varphi}\right)\varphi^{-n}\,
\mathrm{e}^{-\alpha\varphi}+24\,f_0\,h^2\,\mathrm{e}^{\alpha\varphi}\ .
\end{equation}
The present value of the Hubble rate $H_0 \sim 10^{-33}\,\mathrm{eV}$ corresponds to 
$h_0\sim10^{-60}$. If the Hubble rate is given by this value $h_0\sim10^{-60}$, 
we find $n\ll\alpha\varphi_0$. Then we can solve (\ref{e6}) and (\ref{e7}) 
with respect to $\varphi_0$, $v_0$ as follows,
\bea
\label{e8}
\varphi_0&=&\left(\frac{1}{\alpha}\ln\frac{1}{8f_0 h_0\!^2}\right)
\left[1+\frac{1}{n}\,{\cal O}\left(\left(\frac{1}{n}\ln\frac{1}{8f_0h_0\!^2}\right)^{-2}
\right)\right] \\
\label{e9}
v_0&=&\frac{3}{8f_0}\left(\frac{1}{\alpha}\ln\frac{1}{8f_0h_0\!^2}\right)^n
\left[1+\left(\frac{1}{n}\ln\frac{1}{8f_0h_0\!^2}\right)^{-1}
+ {\cal O}\left(\left(\frac{1}{n}\ln\frac{1}{8f_0h_0\!^2}\right)^{-2}\right)\right]\ .
\eea
If we define $V_0$ as $V_0 \equiv v_0\,M_\mathrm{Pl}\!^{4+n}$, $V_0$ has a mass dimension 
$4+n$. Then we may define the mass scale corresponding to $V_0$ by 
$M\equiv V_0^{1/(4+n)} = v_0\!^{1/(4+n)}\,M_\mathrm{Pl}$. Then we find
\begin{equation}
\label{e10}
M\simeq M_\mathrm{Pl}\left[\frac{3}{8f_0}\left(\frac{1}{\alpha}
\ln\frac{1}{8f_0h_0\!^2}\right)^n\right]^{1/(4+n)}\ .
\end{equation}
Then when $n\sim 0$, the magnitude of $M$ is almost equal to that of the Planck scale $M_\mathrm{Pl}$. 
We have found that the scale of $M$, which appear in the potential of the scalar 
field $\phi$, could be the order of the Planck scale although the small scale 
corresponding to the Hubble rate of the present universe. In the following, we investigate how the 
parameter could be restricted by the condition that the correction to the Newton law 
should be small. 

Since the action (\ref{e1}) contains the scalar field $\phi$, if the scalar field couples 
with the matter, the propagation of $\phi$ generates the correction to the Newton law. 
The strength of the coupling could be the order of the inverse of the Planck scale 
$M_\mathrm{Pl}$ and therefore rather small but if the mass of $\phi$ is small, the correction 
could be the same order of the Newtonian force and cannot be neglected. In the bulk of the 
universe, the mass could be the order of the present Hubble rate 
$H_0 \sim 10^{-33}\,\mathrm{eV}$ and very small. For scalar-tensor theories, in order to 
avoid this problem, the so-called Chameleon mechanism is proposed \cite{Chame}, 
where the mass of the 
scalar field $\phi$ could depend on the matter density or scalar curvature. The mass of $\phi$ 
becomes much larger in the solar system or on the earth than in the bulk of the universe and 
the correction to the Newton law could become negligible. 
Here, we propose another kind of Chameleon mechanism, where the coupling of the scalar 
field with the Gauss-Bonnet invariant makes the mass larger in the solar system or on the earth. 

We now investigate how the Chameleon mechanism could work in this model and how restrict 
the magnitude of the mass scale $L = M_\mathrm{Pl}/\alpha$. 
We now check how the correction to the Newton law could be suppressed on the earth. 
Even in the solar system, the correction could be also suppressed. 
On the earth, the equation of the scalar field (\ref{se1}) has the following form:
\begin{equation}
\label{e11}
0=-v_0\left(1+\frac{n}{\alpha\varphi}\right)\varphi^{-n}\,\mathrm{e}^{-\alpha\varphi}
+f_0\,\mathrm{e}^{\alpha\varphi}\,\mathcal{\tilde R}_\mathrm{GB}\!^2\ .
\end{equation}
Since we have $\mathcal{\tilde R}_\mathrm{GB}\!^2\sim10^{-180}$ on the earth, we find 
$n\ll\alpha\varphi_\mathrm{E}$ and we can approximate (\ref{e11}) as follows:
\begin{equation}
\label{e12}
\varphi_\mathrm{E}=\frac{1}{2\alpha}\left\lbrace
\ln\left[\frac{v_0}{f_0\mathcal{\tilde R}_\mathrm{GB}\!^2}\left(\frac{1}{2\alpha}
\ln\frac{v_0}{f_0\mathcal{\tilde R}_\mathrm{GB}\!^2}\right)^{-n}\right] 
 -\ln\left(\frac{1}{2\alpha\varphi_\mathrm{E}}
\ln\frac{v_0}{f_0\mathcal{\tilde R}_\mathrm{GB}\!^2}\right)^{-n}
+ {\cal O}\left(\frac{n}{\alpha\varphi_\mathrm{E}}\right)\right\rbrace\ .
\end{equation}
Since the second term in (\ref{e12}) is much smaller than the first term, 
we obtain the following expression in the mass of the scalar field:
\bea
\label{e13}
m_\phi &\equiv& \left\{\frac{1}{2}\frac{d^2}{d\phi^2}\left(V(\phi)+f(\phi)\,\mathcal{R}_\mathrm{GB}
\right)\right\}^{1/2} \nn
&=& M_\mathrm{Pl}\,\alpha\left\lbrace v_0\left[1+\frac{n}{\alpha\varphi_\mathrm{E}}
+\frac{n\left(n+1\right)}{\alpha^2\varphi_\mathrm{E}\!^2}\right]\varphi_\mathrm{E}\!^{-n}\,
\mathrm{e}^{-\alpha\varphi_\mathrm{E}}+f_0\,\mathrm{e}^{\alpha\varphi_\mathrm{E}}\,
\mathcal{\tilde R}_\mathrm{GB}\!^2\right\rbrace^{1/2} \notag \nn
&\simeq& M_\mathrm{Pl}\,\alpha\left[4f_0v_0\mathcal{\tilde R}_\mathrm{GB}\!^2
\left(\frac{1}{2\alpha}\ln\frac{v_0}{f_0\mathcal{\tilde R}_\mathrm{GB}\!^2}
\right)^{-n}\right]^{1/4}\ .
\eea
By using this equation (\ref{e13}) and (\ref{e9}), we find
\begin{equation}
\label{e14}
m_\phi\simeq M_\mathrm{Pl}\,\alpha\left(\frac{3}{2}\mathcal{\tilde R}_\mathrm{GB}\!^2\right)^{1/4}
\left(\left.\ln\frac{1}{64f_0\!^2h_0\!^4}\right/\left.\ln\frac{v_0}{f_0
\mathcal{\tilde R}_\mathrm{GB}\!^2}\right.\right)^{n/4}\ .
\end{equation}
Since $\left.\ln\frac{1}{64f_0\!^2h_0\!^4}\right/
\left.\ln\frac{v_0}{f_0\mathcal{\tilde R}_\mathrm{GB}\!^2}\right.
\sim {\cal O}(1)$, the $n$-dependence of the mass $m_\phi$ 
is small and we find
\begin{equation}
\label{e15}
m_\phi\sim\alpha\times10^{-18}\,\mathrm{eV}\ .
\end{equation}
In order that the correction to the Newton law could be small, the Compton length of 
the scalar field, which is the inverse of the mass of the scalar field, should 
be smaller than $1\,\mathrm{mm}$ (or $1\,\mu\mathrm{m}$), which corresponds 
to $m_\phi \sim 10^{-3}\,\mathrm{eV}$ (or $m_\phi \sim 1\,\mathrm{eV}$). 
Then we find $\alpha>10^{15}$ (or $\alpha>10^{18}$) and 
$L = M_\mathrm{Pl}/\alpha \sim 10^4\,\mathrm{GeV}$ (or $10\,\mathrm{GeV}$). 
Then we obtain a model, where appearing a very small scale corresponding to the scale of 
the acceleration of the present universe or $H_0$ and the correction to the Newton law 
could be small. More detailed analysis will be given in Sec.\ref{SecIV}. 

In the Chameleon mechanism for the scalar-tensor theory \cite{Chame}, if we observe the scalar mode 
in the vacuum chamber, 
where any matter does not exist inside and therefore the scalar curvature vanishes, the mass of the scalar 
mode becomes very small and the correction to the Newton law could be observed, even on the earth. 
We should note that the scalar curvature and the Ricci tensor vanish in the vacuum but the Riemann 
tensor and therefore the Gauss-Bonnet invariant do not vanish near the earth or in the Solar System, 
even in the vacuum. Therefore in the scenario proposed here, the scalar mode is massive even 
in the vacuum and the correction to the Newton law could be small. 

\section{(In)Stability of de Sitter Solution and Inflation \label{SecIII}}

We now consider the case that there appear another de Sitter space solution beside the solution 
corresponding to the present asymptotically de Sitter space in (\ref{e8}). 
In case such a de Sitter solution exists, the solution may describe the inflation in the early 
universe although the previous solution given by (\ref{e8}) and (\ref{e9}) describes the acceleration 
of the present universe. 

By deleting $h$ from Eqs.(\ref{e6}) and (\ref{e7}), we obtain an equation for $\varphi$: 
\begin{equation}
\label{e16}
\left(1+\frac{n}{\alpha\varphi}\right)\varphi^n=\frac{8}{3}f_0v_0\ .
\end{equation}
Then if $0<n<1$, there are two solutions. One solution, which is denoted by $\varphi_0$, 
corresponds to the previous one in (\ref{e8}) and (\ref{e9}). 
We denote another solution by $\varphi_1$. 
If $n\sim0$, we obtain
\begin{equation}
\label{e17}
\varphi_1\!^n\simeq1+n\ln\varphi_1\ , \quad
\frac{8}{3}f_0v_0\simeq 1+n\left[\ln\left(\frac{1}{\alpha}\ln\frac{1}{8f_0h_0\!^2}\right)
+\left(\ln\frac{1}{8f_0h_0\!^2}\right)^{-1}\right]\ ,
\end{equation}
and therefore
\begin{equation}
\label{e18}
\varphi_1\simeq\frac{1}{\alpha}\left\lbrace\ln\left[\left(\ln\frac{1}{8f_0h_0\!^2}\right)
\ln\left(\ln\frac{1}{8f_0h_0\!^2}\right)\right]+\ln\left[\alpha\varphi_1
\ln\left(\ln\frac{1}{8f_0h_0\!^2}\right)\right]^{-1}
+\left(\ln\frac{1}{8f_0h_0\!^2}\right)^{-1}\right\rbrace^{-1}\ .
\end{equation}
Since the second term in (\ref{e18}) is much smaller than the first term, 
we obtain the following expression of the Hubble rate $H$:
\begin{equation}
\label{e19}
h_1\simeq\left(\frac{1}{8f_0}\right)^{1/2}\exp\left\lbrace 
 -2\ln\left[\left(\ln\frac{1}{8f_0h_0\!^2}\right)
\ln\left(\ln\frac{1}{8f_0h_0\!^2}\right)\right]\right\rbrace^{-1}\ ,
\end{equation}
which gives $h_1\sim0.3$ independent from $\alpha$ and $n$, and therefore $h_1$ could be almost 
the Planck scale. Therefore $h_1$ may correspond to the large Hubble rate corresponding to the 
inflation of the early universe. 
We should also note that $\varphi_1 \ll 1$ although $\varphi_0 \gg 1$. 

Now we consider the case $n\sim 1$. Then we find $n\gg\alpha\varphi_1$ 
and Eq.(\ref{e16}) can be solved with respect to $\varphi_1$ as
\bea
\label{e20}
\varphi_1&\simeq& \left(\frac{8f_0v_0\alpha}{3n}\right)^{-1/(1-n)} \nn
&\simeq& \frac{1}{\alpha}\left\lbrace\frac{1}{n}\left(\ln\frac{1}{8f_0h_0\!^2}\right)^n
\left[1+n\left(\ln\frac{1}{8f_0h_0\!^2}\right)^{-1}\right]\right\rbrace^{-1/(1-n)} \notag \nn
&\simeq& \frac{1}{\alpha}\exp\left\lbrace-1-\frac{n}{1-n}
\left[\ln\left(\ln\frac{1}{8f_0h_0\!^2}\right)+\left(\ln\frac{1}{8f_0h_0\!^2}
\right)^{-1}\right]\right\rbrace\ .
\eea
Then the Hubble rate is given by
\begin{equation}
\label{e21}
h_1\simeq\left(\frac{1}{8f_0}\right)^{1/2}\exp\left\lbrace\frac{1}{2}
+\frac{n}{2\left(1-n\right)}\left[\ln\left(\ln\frac{1}{8f_0h_0\!^2}\right)
+\left(\ln\frac{1}{8f_0h_0\!^2}\right)^{-1}\right]\right\rbrace\ .
\end{equation}
Since $h_0\sim10^{-60}$, $h_1$ diverges when $n\to 1$. Then if we fine-tune the value of $n$, 
we may obtain the large Hubble rate corresponding to the inflation. 

We now investigate the instability of the de Sitter solution. 
We consider the perturbation around the de Sitter solution as
\begin{equation}
\label{e22}
\varphi=\varphi_i+\delta\varphi\ , \quad h=h_i+\delta h \qquad (i=0,1)\ .
\end{equation}
Then Eq.(\ref{e4}) has the following form:
\begin{equation}
\label{e23}
3=v_0\,h_i\!^{-2}\,\varphi_i\!^{-n}\,\mathrm{e}^{-\alpha\varphi_i}
\left[1-2\,\frac{\delta h}{h_i}-\alpha\left(1
+\frac{n}{\alpha\varphi_i}\right)\delta\varphi\right]
+24\,f_0\,\alpha\,h_i\!^2\,\mathrm{e}^{\alpha\varphi_i}\,\delta\varphi'
\end{equation}
and Eq.(\ref{e5}) gives 
\bea
\label{e24}
0&=& \delta\varphi''+3\,\delta\varphi' \vphantom{\frac00}\notag \nn
&\mathrel{\hphantom=}& -v_0\,\alpha\,h_i\!^{-2}\left(1+\frac{n}{\alpha\varphi_i}\right)
\varphi_i\!^{-n}\,\mathrm{e}^{-\alpha\varphi_i}
\left\lbrace1-2\,\frac{\delta h}{h_i}-\alpha\left[1+\frac{n}{\alpha\varphi_i}
+\frac{n}{\alpha^2\varphi_i\!^2}\left(1+\frac{n}{\alpha\varphi_i}
\right)^{-1}\right]\delta\varphi\right\rbrace \notag \nn
&\mathrel{\hphantom=}& +24\,f_0\,\alpha\,h_i\!^2\,\mathrm{e}^{\alpha\varphi_i}
\left(1+\frac{\delta h'}{h_i}+2\,\frac{\delta h}{h_i}+\alpha\,\delta\varphi\right)\ .
\eea
Since $\varphi_i$ and $h_i$ satisfy Eqs.(\ref{e6}) and (\ref{e7}), 
Eq.(\ref{e23}) can be rewritten as
\begin{equation}
\label{e25}
0=2\,\frac{\delta h}{h_i}-\alpha\left(1+\frac{n}{\alpha\varphi_i}
\right)\left(\delta\varphi'-\delta\varphi\right)
\end{equation}
and Eq.(\ref{e24}) has the following form:
\begin{equation}
\label{e26}
0=\delta\varphi''+3\,\delta\varphi'+3\,\alpha\left(1+\frac{n}{\alpha\varphi_i}\right)
\left\lbrace\frac{\delta h'}{h_i}+4\,\frac{\delta h}{h_i}+\alpha\left[2
+\frac{n}{\alpha\varphi_i}+\frac{n}{\alpha^2\varphi_i\!^2}\left(1
+\frac{n}{\alpha\varphi_i}\right)^{-1}\right]\delta\varphi\right\rbrace\ .
\end{equation}
\label{e27}
By using Eqs.(\ref{e25}) and (\ref{e26}) and deleting $\delta h$ and $\delta h'$, 
we obtain the linear differential equation:
\begin{equation}
\label{e28}
0=\delta\varphi''+3\,\delta\varphi'-3\,\alpha^2\left[\frac{n}{\alpha\varphi_i}
+\frac{n\left(n-1\right)}{\alpha^2\varphi_i\!^2}\right]
\left[1+\frac{3}{2}\,\alpha^2\smash{\left(1
+\frac{n}{\alpha\varphi_i}\right)^2}\right]^{-1}\delta\varphi\ .
\end{equation}
We now assume $\delta\varphi \propto \e^{\lambda_i \ln a}$. Then we find
\begin{equation}
\label{e29}
\lambda_i=\lambda_i^\pm \equiv -\frac{3}{2}\pm\frac{3}{2}\left\lbrace1+\frac{4}{3}\,\alpha^2
\left[\frac{n}{\alpha\varphi_i}+\frac{n\left(n-1\right)}{\alpha^2\varphi_i\!^2}\right]
\left[1+\frac{3}{2}\,\alpha^2\smash{\left(1+\frac{n}{\alpha\varphi_i}\right)^2}
\right]^{-1}\right\rbrace^{1/2}\ .
\end{equation}
If the real part of $\lambda_i$ is positive, the perturbation becomes large, which 
tells that the solution is unstable. Since $\lambda_i^-$ in (\ref{e29}) is always negative, 
we now investigate $\lambda_i^+$. 

In the de Sitter solution $\varphi_0$ and $h_0$ corresponding to the present universe 
in (\ref{e8}) or (\ref{e9}), if we assume $\alpha\gg 1$, by using (\ref{e8}), we find
\begin{equation}
\label{e30}
\lambda_0\simeq\frac{2}{3}\,n\left(\ln\frac{1}{8f_0h_0\!^2}\right)^{-1}\ ,
\end{equation}
which is negative and therefore the solution is unstable. 
The instability is, however, very small. In fact, the variation of the scale factor $a$, which 
becomes larger by $\mathrm{e}$-times, is given by
\begin{equation}
\label{e31}
N\equiv\ln\frac{a_\mathrm{F}}{a_\mathrm{I}}\sim\frac{4.1\times10^2}{n}\ ,
\end{equation}
which corresponds to $6\times10^3/n\,\mathrm{Gyr}$ and very large if we use the value of 
the Hubble rate in the present universe. 
Then the de Sitter universe solution corresponding to the present universe could be almost 
stable. 

On the other hand, for another de Sitter solution $\varphi_1$ and $h_1$ in (\ref{e18}), 
when $n\sim0$, we find 
\begin{equation}
\label{e32}
\lambda_1\simeq-\frac{2}{3}\,n\left\lbrace\ln\left[\left(\ln\frac{1}{8f_0h_0\!^2}\right)
\ln\left(\ln\frac{1}{8f_0h_0\!^2}\right)\right]\right\rbrace\left\lbrace
\ln\left[\left(\ln\frac{1}{8f_0h_0\!^2}\right)\ln\left(\ln\frac{1}{8f_0h_0\!^2}
\right)\right]-1\right\rbrace\ .
\end{equation}
When $n\sim 1$, since $n\gg\alpha\varphi_1$, we find 
\begin{equation}
\label{e33}
\lambda_1\simeq-\frac{2}{3}\left(1-n\right)\ .
\end{equation}
In both of cases $n\sim 0$ and $n\sim 1$, $\lambda_1$ is always positive and 
de Sitter solution is always stable.

Since the Hubble rate $h_1$ is large and the Planck scale in the solution (\ref{e18}), 
the solution might describe the inflation. As we have seen in (\ref{e32}) or (\ref{e33}), 
however, the solution is stable and therefore the inflation could be eternal. 
In order to make an exit 
from the inflation, we may add the following term with scalar field $\zeta$ coupled with the 
scalar curvature, to the action (\ref{e1}):
\be
\label{e34}
\Delta S = \int d^4 x \sqrt{-g}\left\{ - \frac{1}{2}\partial_\mu \zeta \partial^\mu \zeta 
+ c \zeta \mathcal{R}\right\}\ .
\ee
Here $c$ is a constant. By the variation with respect $\zeta$, we obtain
\be
\label{e35}
\Box \zeta + c \mathcal{R} = 0 \ .
\ee
or $\zeta = - c \Box^{-1} \mathcal{R}$. Therefore the action (\ref{e34}) is equivalent to the non-local 
action: 
\be
\label{e36}
\Delta S = \int d^4 x \sqrt{-g}\left\{ - \frac{c^2}{2} \mathcal{R}\Box^{-1} \mathcal{R}\right\}\ .
\ee
Note that similar non-local action for GB invariant leads to one (several)
scalar-Gauss-Bonnet gravity (see \cite{Capozziello:2008gu}). 
In the FRW universe, if we assume $\zeta$ only depends on $t$ or $\ln a$, Eq.(\ref{e35}) 
has the following form:
\be
\label{e37}
0= \zeta'' + \left(3 + \frac{h'}{h}\right) \zeta' - 6c \left(\frac{h'}{h} + 2 \right)\ .
\ee
We assume that $\Delta S$ could be very small at the beginning of the universe. 
Then we now consider the time-development of $\zeta$ in the de Sitter background. 
In the de Sitter space, where $\mathcal{R}=12H_1^2$ $\left(H_1 = h_1 M_\mathrm{Pl}\right)$, 
$\zeta$ can be solved as
\be
\label{e38}
\zeta = 4c \ln a - \zeta_1 a^{-3} + \zeta_2\ .
\ee
Here $\zeta_1$ and $\zeta_2$ are constants of the integration. 
Eq.(\ref{e38}) tells that $\zeta$ becomes larger by the expansion of the universe. 
Then the magnitude of $\Delta S$ in (\ref{e34}) could become larger and dominant. When 
$\Delta S$ is dominant, the equation corresponding to the FRW equation has the following form:
\be
\label{e39}
\frac{6}{\kappa^2}= \left(\zeta'\right)^2 - 12c \zeta' - 12c\zeta \ .
\ee
If we change the variable $\zeta$ to $\eta$ as
\be
\label{e40}
\zeta= - 3c + \frac{1}{2c\kappa^2} + \frac{1}{12c}\left(\eta - 6c \right)^2\ ,
\ee
Eq.(\ref{e39}) can be rewritten as
\be
\label{e41}
\left(\eta - 6c \right) \eta' = 6 c\eta\ ,
\ee
which can be solved and we find
\be
\label{e42}
\frac{\eta}{6c} - \ln \frac{\eta}{\eta_0} = \ln a\ \mbox{or}\ \eta=0\ .
\ee
Here $\eta_0$ is a constant of the integration. 
Eq.(\ref{e42}) has two branches, that is, a value of $\ln a$ corresponds to two values of $\eta$. 
If we assume the first term in the solution in (\ref{e42}) dominates for large $a$, 
by using (\ref{e40}), we find
\be
\label{e43}
\eta \sim 6c\ln a\ .
\ee
Then by solving (\ref{e37}) with respect to $h$, we find $h\propto a^{-3}$ or $a\propto t^{1/3}$. 
On the other hand, if the second term in (\ref{e42}) dominates, we find $\eta\to 0$, 
which corresponds to the second solution $\eta=0$ in (\ref{e42}). In this case, 
we find $h\sim a^{-2}$ or $a\propto t^{1/2}$. 
In any case, the Hubble rate $H$ is proportional to $1/t$ and therefore $\mathcal{R}$ 
is proportional to $1/t^2$. 
Then the curvature becomes smaller and the de Sitter phase or inflation will stop. 
Typically the de Sitter phase could stop when the order of magnitude of $\Delta S$ 
becomes that of the Einstein-Hilbert action, that is, typically 
$\mathcal{R}/2\kappa^2 \sim c\zeta \mathcal{R}$. 
By using (\ref{e38}), we find $c \zeta \sim 1/\kappa^2 \sim M_\mathrm{Pl}\!^2$. 
Then from (\ref{e38}), we find that the sufficiently large $e$-folding as $50$-$60$ 
could be obtained if $M_\mathrm{Pl}/c \sim 20$. 
When $\Delta S$ dominates, $\mathcal{R}$ behaves as $1/t^2$ and therefore 
$\mathcal{R}_\mathrm{GB} \sim \mathcal{R}^{(0)}_\mathrm{GB}/t^4$ with a constant 
$\mathcal{R}^{(0)}_\mathrm{GB}$.
The value of the scalar field $\phi$ is given by the minimum of 
the effective potential $V(\phi) + f(\phi)\,\mathcal{R}_\mathrm{GB}$, that is
\be
\label{eff1}
0 = V'(\phi) + f'(\phi)\,\mathcal{R}_\mathrm{GB}\!^2 
\sim V(\phi) + \frac{f(\phi)\,R^{(0)}_\mathrm{GB}\!^2}{t^4}\ ,
\ee
which could be solved, by using (\ref{e2}), as
\be
\label{eff2}
\phi \sim \frac{4M_\mathrm{Pl}}{\alpha} \ln \frac{t}{M_\mathrm{Pl}}\ .
\ee
Then the value of $\phi$ becomes larger and therefore the value of the (effective) 
potential becomes smaller, which may generate 
the reheating and there could occur the particle production by the oscillation 
of $\phi$ around the minimum of the effective potential. 
Then universe may go into the matter dominated phase. Even in the 
matter dominated phase, the scalar curvature $\mathcal{R}$ is proportional to $1/t^2$ and therefore 
the value of the scalar field $\phi$ becomes larger and larger and finally 
the de Sitter solution with the small Hubble rate corresponding to (\ref{e8}) could 
be realized and the accelerating expansion of the present universe could occur. 
For more quantitative arguments, we may need numerical calculations, which could be a future work. 

Let the value of $\zeta=\zeta_0$ in the present universe. 
By writing $\zeta$ as $\zeta = \zeta_0 + \delta \zeta$, the action in the present 
universe has the following form: 
\be
\label{e44}
S_\mathrm{total} = \int d^4 x \sqrt{-g}\left\{\left(\frac{1}{2\kappa^2} + c\zeta_0\right)\mathcal{R} 
 - \frac{1}{2}\partial_\mu \delta\zeta \partial^\mu \delta\zeta 
+ c \delta\zeta \mathcal{R} + \mbox{$\phi$ and Gauss-Bonnet terms}\right\}\ .
\ee
Then we may identify the present gravitational coupling $\kappa_\mathrm{present}$ as
\be
\label{e45}
\frac{1}{2\kappa_\mathrm{present}^2}=\frac{1}{2\kappa^2} + c\zeta_0\ .
\ee
If we choose the parameter as before $1/c\kappa_\mathrm{present} \sim {\cal O}(10)$, 
the time-development of the scalar field $\zeta$ is very small since the value 
of ${\cal O}(10))$ of the $e$-folding corresponding to 
about one hundred of billion years. 
Then different from the case of the time-development of $\zeta$ in the inflation of the 
early universe, the time-development of $\zeta$ in the present universe is very slow and 
asymptotic de Sitter universe will continue about one hundred of billion years. 

\section{E\" otv\"os Experiment and Correction to Newton Law \label{SecIV}}

We now investigate if the model could satisfy the constraint from 
E\" otv \"os experiment and the Newton law. 
Although the correction coming from the scalar field could be small since 
the mass $m$ of the scalar field can be large, the point source shifts the magnitude of 
the Gauss-Bonnet invariant. Since the scalar field $\phi$ coupled with the Gauss-Bonnet 
invariant, the scalar field $\phi$ itself could be shifted and another correction to the Newton law 
could be generated. 

First we mention about E\" otv\"os experiment. 
Since the Gauss-Bonnet invariant $\mathcal{R}_\mathrm{GB}$ slowly changes with the radius coordinate 
$r$ compared with the mass $m$ of the scalar field $\phi$, 
$\left|d\ln \mathcal{R}_\mathrm{GB}\!^2/dr\right| \ll m$, we can neglect the term coming kinetic term 
in the scalar field equation (\ref{se1}) and the radius $r$ dependence of $\phi$ could be 
determined by the minimum of the effective potential 
$V_\mathrm{eff} = V(\phi) + f(\phi)\mathcal{R}_\mathrm{GB}(r)^2$: 
\be
\label{A0}
0 = V'\left(\phi(r)\right) + f'\left(\phi(r)\right) \mathcal{R}_{\mathrm{GB}\oplus}(r)^2\ .
\ee
If we assume the strength of the coupling of $\phi$ with matter of kind $i$ 
is given by $\beta_i / M_\mathrm{Pl}$, the magnitude of the force $F_i$ coming from 
the scalar field could be given by
\be
\label{A1}
F_i \sim \frac{\beta_i M_i}{M_\mathrm{Pl}}
\left. \frac{d\phi_\mathrm{min}}{dr}\right|_{r=R_\oplus}\ .
\ee
We now define the E\" otv\"os parameter $\eta$ by
\be
\label{A2}
\eta \equiv 2 \frac{\left|a_1 - a_2 \right|}{a_1 + a_2} 
\sim \frac{8\pi\left| \beta_1 - \beta_2\right| 
M_\mathrm{Pl} R_\oplus\!^2}{M_\oplus} 
\left. \frac{d\phi_\mathrm{min}}{dr} \right|_{r=R_\oplus}
\ee
Here $M_\oplus$ is the mass of the earth. 

Before investigating how the model (\ref{e2}) can satisfy the constraint for the E\" otv\"os 
parameter $\eta$ from the experiment, we consider how we should investigate the correction to 
the Newton law between the two massive point particles on the earth. 
The correction can be found by considering the fluctuation $\delta \phi$ 
of around the background value $\phi_\oplus$ of the scalar field $\phi$:
\be
\label{A3}
\phi = \phi_\oplus + \delta \phi\ .
\ee
Here as in (\ref{A0}), $\phi_\oplus$ could be determined by 
\be
\label{A4}
0 = V'\left(\phi_\oplus\right) + f'\left(\phi_\oplus\right) \mathcal{R}_{\mathrm{GB}\oplus}^2\ .
\ee
We assume the distance between two particles is sufficiently short compared with the length scale 
of the earth. Then $\phi_\oplus$ could be regarded to be constant. 
Then $\mathcal{R}_\mathrm{GB}\!^2 $ could be given by the of contributions from the earth 
and one of point particles: 
\be
\label{A5}
\mathcal{R}_\mathrm{GB}\!^2 = \mathcal{R}_{\mathrm{GB}\oplus}^2 
+ \frac{3M_i^2}{4\pi^2 M_\mathrm{Pl}\!^4 r^6}\ .
\ee
Here $r$ is the distance from the point particle. Then from (\ref{se1}), we find
\be
\label{A6}
\left(\nabla^2 - m_\oplus\!^2 \right) \delta\phi 
= \frac{3 f'\left(\phi_\oplus\right) M_i^2}{4\pi^2 M_\mathrm{Pl}\!^4 r^6} 
+ \frac{\beta_i M_i}{M_\mathrm{Pl}}\delta^{(3)}\left(\bm{r}\right)\ .
\ee
Here the mass $m_\oplus\!^2$ of the scalar field on the earth is given by
\be
\label{A7}
m_\oplus\!^2 \equiv V''\left(\phi_\oplus\right) 
+ f''\left(\phi_\oplus\right) \mathcal{R}_{\mathrm{GB}\oplus}^2\ .
\ee
Then by assuming $m_\oplus\!^2$ is large enough, that is, 
$\left|d\ln {\cal R}_\mathrm{GB}\!^2/dr\right| \ll m_\oplus$, 
we can solve Eq.(\ref{A6}) with respect 
to $\delta\phi$ as follows, 
\bea
\label{A8}
\delta \phi(r) &=& \int d^3\bm{r} \left[ 
\frac{3 f'\left(\phi_\oplus\right) M_i^2}{4\pi^2 M_\mathrm{Pl}\!^4 r^6} 
+ \frac{\beta_i M_i}{M_\mathrm{Pl}} \delta^{(3)}\left(\bm{r}\right) 
\right]
\frac{\e^{-m_\oplus\left|\bm{r} - \bm{r}'\right|}}{4\pi \left|\bm{r} - \bm{r}'\right|} \nn
&\sim & \frac{\beta_i M_i}{M_\mathrm{Pl}} \delta^{(3)}\left(\bm{r}\right) \frac{\e^{-m_\oplus r}}{4\pi r}
+ \frac{3 f'\left(\phi_\oplus\right) M_i^2}{4\pi^2 M_\mathrm{Pl}\!^4 r^6}
\int d^3\bm{r} \frac{\e^{-m_\oplus\left|\bm{r} - \bm{r}'\right|}}{4\pi \left|\bm{r} - \bm{r}'\right|} \nn
& = & \frac{\beta_i M_i}{M_\mathrm{Pl}} \delta^{(3)}\left(\bm{r}\right) \frac{\e^{-m_\oplus r}}{4\pi r}
+ \frac{3 f'\left(\phi_\oplus\right) M_i^2}{4\pi^2 M_\mathrm{Pl}\!^4 m_\oplus\!^2 r^6}\ .
\eea
Then the potential between the two particles could be given by
\be
\label{A9}
\mathcal{V} \sim \frac{\beta_1 \beta_2 M_1 M_2}{4\pi M_\mathrm{Pl}\!^2}
\frac{\e^{-m_\oplus r}}{r} 
+ \frac{3f'\left(\phi_\oplus\right) M_1 M_2 \left(M_1 + M_2\right)}
{8\pi^2 M_\mathrm{Pl}\!^5 m_\oplus\!^2 r^6}\ .
\ee
Then the ratio between the correction and the Newton force could be given by
\be
\label{A10}
\alpha \sim 2 \beta_1 \beta_2 \e^{-m_\oplus r} 
+ \frac{3f'\left(\phi_\oplus\right) \left(M_1 + M_2\right)}
{\pi M_\mathrm{Pl}\!^3 m_\oplus\!^2 r^5}\ .
\ee

For our model (\ref{e2}), we find
\be
\label{A11}
\phi_\oplus \sim \frac{L}{2}
\ln \left[ \frac{M^4}{f_0 \mathcal{R}_{\mathrm{GB} \oplus}^2}
\left(\frac{L}{2M}\ln \frac{M^4}{f_0 \mathcal{R}_{\mathrm{GB} \oplus}^2}
\right)^{-n}\right]\ .
\ee
Here $L\equiv M_\mathrm{Pl}/\alpha$ and the mass could be given by 
\be
\label{A12}
m_\oplus ^2 = \frac{M^{4+n}}{L^2}\left[ 1 + \frac{2nL}{\phi_\oplus} 
+ \frac{n(n+1) L^2}{\phi_\oplus\!^2} \right] \phi_\oplus\!^{-n} \e^{-\phi_\oplus /L} 
+ \frac{f_0}{L^2}\e^{\phi_\oplus/L} \mathcal{R}_\mathrm{GB}\!^2
\sim \frac{2M^2}{L^2} \left[ f_0 \mathcal{R}_\mathrm{GB}\!^2 
\left( \frac{L}{2M} \ln \frac{M^4}{f_0 \mathcal{R}_\mathrm{GB}\!^2} \right)^{-n}\right]^{1/2}\ ,
\ee
which can be rewritten, by using (\ref{e10}), as 
\be
\label{A13} 
m_\oplus\!^2 \sim \frac{M_\mathrm{pl}}{L}\left[ \frac{3}{2} 
\mathcal{R}_\mathrm{GB}\!^2 \left(\ln \frac{M_\mathrm{Pl}\!^4}{64 f_0^2 H_0^4}
\left/ \ln\frac{M^4}{f_0 \mathcal{R}_\mathrm{GB}} \right. \right)^n \right]^{1/4} 
\sim \frac{M_\mathrm{pl}}{L}\left( \frac{3}{2} 
\mathcal{R}_\mathrm{GB}\!^2 \right)^{1/4}\ .
\ee
By assuming that the Gauss-Bonnet invariant could be given by the Schwarzschild metric 
of the earth, 
\be
\label{A16}
\mathcal{R}_\mathrm{GB}\!^2 \sim \frac{3M_\oplus\!^2}{4\pi^2 M_\mathrm{Pl}\!^4 r^6} \ ,
\ee
the E\" otv\"os parameter $\eta$ in (\ref{A2}) could be given by
\be
\label{A17}
\eta \sim \frac{4\pi \left| \beta_1 - \beta_2 \right| M_\mathrm{Pl} R_\oplus L}{M_\oplus}\ .
\ee
In order to satisfy the constraint $\eta < 10^{-13}$ from the experiment, since 
$M_\mathrm{Pl} \sim 10^{27}\mathrm{eV}$, $M_\oplus \sim 10^{60} \mathrm{eV}$, 
$R_\oplus \sim 10^{13}\mathrm{eV}^{-1}$, and $\left| \beta_1 - \beta_2 \right| \lesssim 10^{-4}$, 
we find
\be
\label{A18}
L \lesssim 10\mathrm{GeV}\ .
\ee
On the other hand, in the ratio between the correction and the Newton force in (\ref{A10}), 
the first term could be estimated by requiring $m_\oplus \gtrsim 10^{-2}\mathrm{eV}$ 
as in (\ref{e15}) and using $\mathcal{R}_{\mathrm{GB}\oplus}^2 \sim 10^{-71}\mathrm{eV}^4$, 
we find 
\be
\label{A19}
L \lesssim 10^3\mathrm{GeV} \ ,
\ee
which corresponds to the results after (\ref{e15}). 
When the second term in (\ref{A10}) dominates, we obtain
\be
\label{A20}
\alpha \sim \frac{LM^2\left(M_1 + M_2\right)}{\pi M_\mathrm{Pl}\!^5 r^5 
\mathcal{R}_{\mathrm{GB}\oplus}^2}\left[ 6f_0 
\left(\frac{L}{2M}\ln \frac{M^4}{f_0 \mathcal{R}_{\mathrm{GB}\oplus}^2}\right)^{-n}\right]^{1/2}\ .
\ee
In order to satisfy the constraint $\alpha\lesssim 10^{-2}$ from the experiments, 
since $M_i \sim 10^{34}\mathrm{eV}$ and $r\sim 10^4 \mathrm{eV}^{-1}$ in the experiments, 
we find the strongest constraint:
\be
\label{A21}
L \lesssim 10^{-6}\mathrm{eV} \quad \mbox{when}\ n\sim 0\ .
\ee
The obtained scale $L$ is much larger than the scale of the Hubble rate in the present universe, 
$H_0 \sim 10^{-33}\mathrm{eV}$ but much smaller than the Planck scale or the scale of 
the particle physics. 

We should note that there is one more problem. The effective potential $V_\mathrm{eff}$ 
could be estimated to be 
\be
\label{A14}
V_\mathrm{eff}\left(\phi_\oplus\right) \sim 
M_\mathrm{Pl}\!^2\left[ \frac{3}{2} 
\mathcal{R}_\mathrm{GB}\!^2 \left(\ln \frac{M_\mathrm{Pl}\!^4}{64 f_0^2 H_0^4}
\left/ \ln\frac{M^4}{f_0 \mathcal{R}_\mathrm{GB}} \right. \right)^n \right]^{1/2} 
\sim M_\mathrm{pl}^2 \left( \frac{3}{2} 
\mathcal{R}_\mathrm{GB}\!^2 \right)^{1/2}\ .
\ee
The effective potential $V_\mathrm{eff}$ plays the role of the cosmological constant, this gives a 
contribution to the Gauss-Bonnet invariant in addition to the contribution from the Schwarzschild 
metric of the earth. Then when we include the contribution, Eq.(\ref{A16}) is replaced by
\be
\label{A15}
\mathcal{R}_\mathrm{GB}\!^2 \sim \frac{3M_\oplus\!^2}{4\pi^2 M_\mathrm{Pl}\!^4 r^6} 
+ \frac{8}{3} \frac{V_\mathrm{eff}\left(\phi_\oplus\right)}{M_\mathrm{Pl}\!^4}
\sim \frac{3M_\oplus\!^2}{4\pi^2 M_\mathrm{Pl}\!^4 r^6} + 4 \mathcal{R}_\mathrm{GB}\!^2 \ ,
\ee
which tells $\mathcal{R}_\mathrm{GB}\!^2$ could be negative. When 
$\mathcal{R}_\mathrm{GB}\!^2$ is negative, Eq.(\ref{A4}) has no solution. 
The problem could occur by the behavior of $f(\phi$ in our model (\ref{e2}). 
In order to have a solution in (\ref{se1b}) or (\ref{e7}), $f(\phi)$ must be large since 
the curvature in the bulk space is very small. In the bulk, the value of $\phi$ is large. 
In our choice of $f(\phi)$ in (\ref{e2}), 
$f(\phi)$ is still large even on the earth, where $\phi$ is smaller than the value in the bulk, 
and we obtained (\ref{A15}). This problem is 
also related with the strong constraint in (\ref{A21}). 
In order to solve this problem, we may choose $f \propto \e^{\alpha^{(n)}\phi^n/M_\mathrm{Pl}}$ 
with a positive constant $\alpha^{(n)}$. 
In this choice, $f(\phi)$ could become large more rapidly for large $\phi$, which could correspond 
to the bulk space and relatively, $f(\phi)$ could become much smaller for the curvature on the 
earth. 

\section{Summary \label{SecV}}

We have proposed an extension of the Chameleon mechanism where the scalar mode 
in the scalar-Einstein-Gauss-Bonnet gravity model becomes massive due 
to the coupling with the Gauss-Bonnet term and therefore the correction to the 
Newton law could be small. 
Since the Gauss-Bonnet invariant does not vanish 
near the earth or in the Solar System, even in the vacuum, the scalar mode is massive even 
in the vacuum and the correction to the Newton law could be small. 
We have also discussed about the possibility that the model could describe simultaneously 
the inflation in the early universe, in addition to the current accelerated expansion. 

We have also discussed about the possibility that the model could describe simultaneously 
the inflation in the early universe, in addition to the current accelerated expansion. 
If we choose $0<n<1$ in the potential of (\ref{e2}), there appear two de Sitter space 
solutions. In one of the solution, the Hubble rate is the order of the Planck scale and 
very large. The solution with the large Hubble rate could be identified with the inflation. 
In order to make an exit from the inflation, 
we may add a small term given by another scalar field coupled with the scalar curvature, 
which could be equivalent to the non-local action and generates the instability of 
the de Sitter solution. The added term is relevant only in the 
epoch of the inflation but irrelevant to the present accelerating universe. 
The oscillation of the scalar field may generate the reheating of the universe. 
In string-inspired gravity with higher order terms there are $R^3$ and $R^4$ 
terms, etc. coupled with dilaton and/or with other scalars.
It would be interesting to estimate the role of such higher-order terms to
chameleon mechanism.

\section*{Acknowledgments}

We are grateful to S D Odintsov for very helpful discussions.
The work by S.N. is supported in part by the Ministry of Education,
Science, Sports and Culture of Japan under grant no.18549001 and Global
COE Program of Nagoya University provided by the Japan Society
for the Promotion of Science (G07).

\end{document}